\documentclass[twocolumn, showpacs, prl, floatfix]{revtex4}

\usepackage{graphicx}

\usepackage{dcolumn}

\usepackage{bm}

\DeclareGraphicsExtensions{.eps,.png}

\begin{document}

\title{Valley polarization and susceptibility of composite fermions around \(\nu = \frac{3}{2}\)}

\date{\today}

\author{N.C.\ Bishop}
\author{M.\ Padmanabhan}
\author{K.\ Vakili}
\author{Y.P.\ Shkolnikov\footnote{currently at Exponent, Inc., Philadelphia, PA}}
\author{E.P.\ De Poortere\footnote{currently at Columbia University, New York, NY}}
\author{M.\ Shayegan}

\affiliation{Department of Electrical Engineering, Princeton University, Princeton, NJ 08544 USA}

\begin{abstract}

We report magnetotransport measurements of fractional quantum Hall states in an AlAs quantum well around Landau level filling factor \(\nu = \frac{3}{2}\), demonstrating that the quasiparticles are composite Fermions (CFs) with a valley degree of freedom.  By monitoring the valley level crossings for these states as a function of applied symmetry-breaking strain, we determine the CF valley susceptibility and polarization.  The data can be explained well by a simple Landau level fan diagram for CFs, and are in nearly quantitative agreement with the results reported for CF spin polarization.

\end{abstract}

\pacs{71.10.Pm, 71.70.Fk, 73.43.-f}

\maketitle

Currently there is considerable interest in quantum Hall effect in multi-valley two-dimensional electron systems (2DESs) \cite{tokeCM2007}. The interest partly stems from the recent magnetotransport results in graphene \cite{novselovNature2005,zhangNature2005} where there is a two-fold valley degeneracy.  2DESs confined to certain semiconductors such as Si or AlAs, too, possess a valley degree of freedom, and the role of valley degeneracy in the interaction induced integer and fractional quantum Hall effects (IQHE and FQHE) was indeed experimentally studied recently in these systems \cite{poortereAPL2002, shkolnikovPRL2005, shayeganPSSb2006, laiPRL2004}.  Here we report measurements of FQHE states in a 2DES confined to an AlAs quantum well. This system has the rather unique property that its valley degeneracy can be controlled via the application of symmetry-breaking in-plane strain \cite{shkolnikovPRL2005, shayeganPSSb2006}.  We focus on the FQHE states around the Landau filling factor \(\nu = \frac{3}{2}\) and measure their strengths as a function of strain.  The data can be well described in a composite Fermion (CF) picture \cite{jainPRL1989,hlrPRB,perspectivesQHE} where the CFs have an additional (valley) degree of freedom. From the experimental data we deduce the ``valley susceptibility" \cite{gunawanPRL2006}, defined as the change in valley polarization as a function of strain, for both electrons and CFs. The results reveal that this susceptibility is significantly enhanced for the CFs over the band value, and shows a strong dependence on the filling factor \(2 > \nu > 1\).  Comparison of our data with the results of {\it spin} polarization measurements \cite{duPRL1995} and calculations \cite{parkPRL1998} for CFs around \(\nu = \frac{3}{2}\) reveals the remarkable similarity of the polarization phase diagram for CFs in both systems, highlighting the analogy between spin and valley degrees of freedom.

\begin{figure}
\centering
\includegraphics[width=80mm]{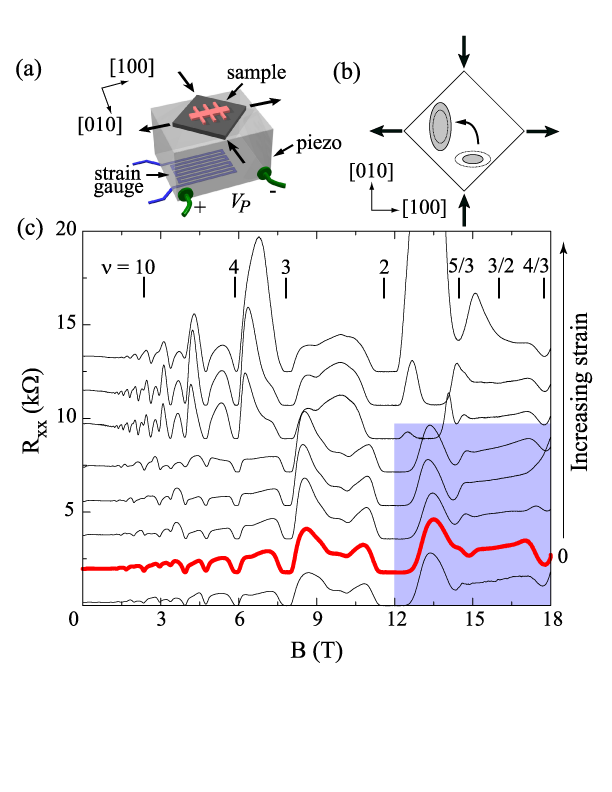}
\caption{(color online) (a) Schematic of the sample, piezoelectric stack, and strain gauge.  (b) Diagram showing transfer of electrons from the [100] to the [010] valley as a function of applied strain, at constant density.  (c) Magnetoresistance traces at a density of \(5.7 \times 10^{11}\) cm\(^{-2}\) for different values of applied strain at \(T \simeq 50\)mK.  The applied strain for each trace, from bottom to top, is \(-0.2, 0.0, 0.3, 0.6, 1.1, 3.4, 3.6, 4.1 \times 10^{-4}\), with zero strain shown in red.  The traces are vertically offset for clarity.  The area in the shaded box is magnified in Fig. 2.}
\end{figure}

We performed experiments on two samples, each containing a 2DES confined to a layer of AlAs, one 11nm and the other 15nm thick, and modulation-doped with Si.  The samples were grown by molecular beam epitaxy on semi-insulating (001) GaAs substrates.  Results from both samples were similar; we focus here on the 11nm sample.  In these samples, electrons occupy two in-plane conduction-band minima (valleys) at the X point in the Brillouin zone, with elliptical Fermi contours as schematically shown in Fig. 1(b) \cite{shayeganPSSb2006, poortereAPL2002}. We refer to these valleys by the direction of their major axis ([100] and [010]).  We deposited AuGeNi contacts on a lithographically defined Hall-bar mesa, aligned along the [100] crystal direction, as shown in Fig. 1(a).  Mounting the sample on a piezoelectric stack (Fig. 1(a)) with its [100] axis along the stack's poling direction allows us to apply controllable, symmetry-breaking strain, by changing the voltage (\(V_{P}\)) on the stack \cite{shayeganAPL2003}. When positive (negative) \(V_{P}\) is applied, e.g., the piezo stack expands (shrinks) along its poling direction and shrinks (expands) in the perpendicular direction, straining the sample.  We define this strain as \(\epsilon = \epsilon_{[100]} - \epsilon_{[010]}\), where \(\epsilon_{[100]}\) and \(\epsilon_{[010]}\) are the strains along the [100] and [010] directions, respectively.  Such strain transfers electrons from the [100] valley to the [010] valley while the total density remains constant, as schematically shown in Fig. 1(a).  The applied strain was measured using a strain gauge glued to the other side of the piezo (see Fig. 1(a)) \cite{shayeganAPL2003}.  We present data taken at a density of \(5.7 \times 10^{11}\) cm\(^{-2}\)  with a mobility of \(150,000\) cm\(^{2}\)/Vs.  All measurements were done in a top-loading dilution refrigerator with a base temperature of \(\simeq 50\)mK in an 18T superconducting magnet.

\begin{figure}
\centering
\includegraphics[width=85mm]{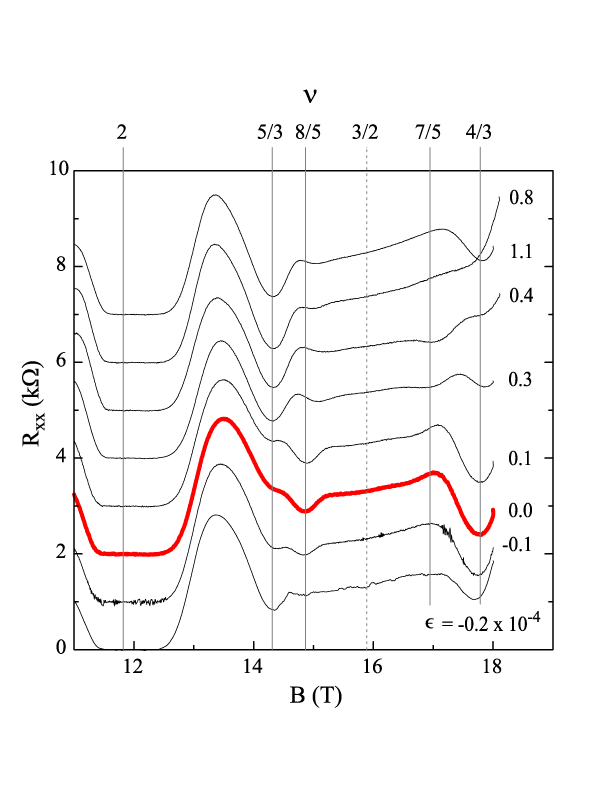}
\caption{High-field magnetoresistance data for different values of applied strain at \(T \simeq 50\)mK.  The thick trace shown in red denotes minimum strain.  Solid vertical lines mark FQHE minima; positions are calculated from the periodicity of the low field Shubnikov-de Haas oscillations.  The numbers shown on the right side give the value of applied strain for each trace.}
\end{figure}

Figure 1(c) gives an overview of our magnetoresistance data.  The strengths of both the IQHE and FQHE states clearly vary with applied strain.  Figure 2 shows the FQHE states around \(\nu = \frac{3}{2}\) in more detail, with additional intermediate strain values included.  Notice that all the FQHE states shown in Fig. 2 disappear or weaken, and reappear as a function of strain.  When the two valleys are balanced ({\it i.e.}, are equally occupied) the \(\nu = \frac{5}{3}\) minimum is very weak, but the \(\nu = \frac{4}{3}\) and \(\frac{8}{5}\) minima are strong.  Away from balance, the \(\nu = \frac{5}{3}\) minimum is strong, but the \(\nu = \frac{4}{3}\) and \(\frac{8}{5}\) minima become weak and then strong again as the magnitude of strain increases.  

\begin{figure}
\centering
\includegraphics[width=85mm]{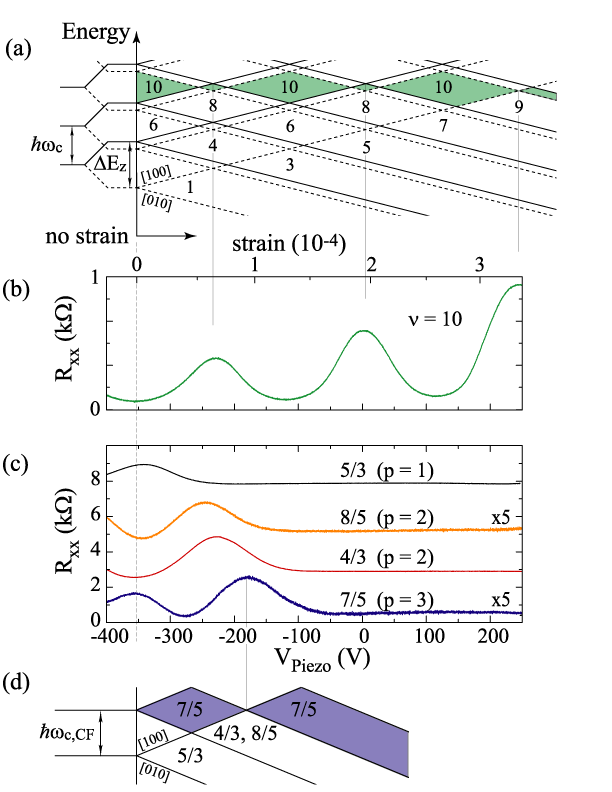}
\caption{(a) Energy fan diagram for electrons showing Landau level crossings as a function of strain for constant magnetic field.  (b) Piezoresistance trace taken at \(\nu = 10\) at \(T \approx 50\)mK.  Note that the small diamonds shown in part (a) are not resolved in this measurement.  (c) Piezoresistance traces taken at FQHE minima at \(T \approx 50\)mK.  Traces have been offset for clarity, and data for \(\nu = \frac{8}{5}\) and \(\frac{7}{5}\) have been magnified vertically to improve contrast.  (d) Proposed energy fan diagram for CF Landau levels.}
\end{figure}

We first discuss the simpler case of the IQHE data.  A magnetic field (\(B\)) applied perpendicular to a 2DES quantizes the allowed energies into Landau levels (LLs), separated by the cyclotron energy \(\hbar \omega_{c}\).  Because the electron spins also interact with the magnetic field, each level is further split into spin up and spin down levels, separated by the Zeeman energy \(\Delta E_{Z} = g^*\mu_{B} \cdot B\), where \(g^*\) is the effective g-factor and \(\mu_{B}\) is the Bohr magneton \cite{footnote1}.  By applying strain to the system, each of the spin-split LLs splits again into [100] and [010] valley levels, separated by the valley splitting energy \(\Delta E_{v} = E^{*}_{2} \cdot \epsilon \), where \(E^{*}_{2}\) is the effective conduction band deformation potential \cite{footnote2}.  This is illustrated in Fig. 3(a), showing the LLs behavior at constant \(B\) under applied strain \cite{footnote2}.  As clearly seen, for certain strain values, LLs with opposite valley polarization coincide, and this coincidence happens periodically as the strain is increased.  It is well known that in a transport measurement, resistance is high when two energy levels are coincident at the Fermi energy, and decreases as the system moves away from coincidence \cite{fangPR1968}.  Figure 3(b) shows the measured resistance as a function of strain at the \(\nu = 10\) IQHE state, which is consistent with the behavior predicted by the simple LL picture \cite{footnote3}.

Before analyzing the CF coincidence data, some discussion of the details of CF states would be helpful.  The CFs around \(\nu = \frac{3}{2}\) are holes in the second LL, that is, the carrier component of the flux-carrier composite particle is a hole, even though at \(B = 0\) the carriers are electrons.  In contrast to CFs around \(\nu = \frac{1}{2}\), the density of \(\nu = \frac{3}{2}\) CFs (\(n_{CF}\)) is not equal to the density of electrons (\(n_{el}\)), and increases with decreasing \(\nu\).  At \(\nu = \frac{3}{2}\), \(n_{CF} = n_{el} / 3\).  Since each CF carries two flux quanta (\(2h / e\)) of the gauge field, which opposes the applied field, the CFs experience no net magnetic field at \(\nu = \frac{3}{2}\).  Thanks to the variation of \(n_{CF}\) with magnetic field, the CFs feel an effective magnetic field \(B_{eff} = 3(B - B_{3/2})\), where \(B_{3/2}\) is the magnetic field at \(\nu = \frac{3}{2}\) \cite{hlrPRB, willettSST1997}. This is because each additional flux quantum of external field creates another hole, which carries two flux quanta of gauge field.  The CFs form LLs under this \(B_{eff}\), and thus the FQHE states around \(\nu = \frac{3}{2}\), can be thought of as simply the IQHE of the CFs, and the CF filling factor \(p\)  is related to the electron filling factor \(\nu\) by \(\nu = 2 - p / (2p \pm 1)\).

With this picture in mind, we construct a LL fan diagram (Fig. 3(d)) for the CF states.  This diagram is qualitiatively consistent with the CF piezoresistance data shown in Fig. 3(c) \cite{footnote4}.  For example, since \(\nu = \frac{5}{3}\) corresponds to \(p = 1\), one can see from the fan diagram that there should be no gap for balanced valleys, and that the gap should widen and eventually saturate as the applied strain increases.  The piezoresistance trace for \(\nu = \frac{5}{3}\) is consistent with this, showing high resistance (very small or no gap) at balance, then falling to a low resistance which saturates at large applied strain.  The \(p = 2,3\) traces likewise show consistent behavior: the \(p = 2\) states, at \(\nu = \frac{4}{3}\) and \(\frac{8}{5}\), exhibit low resistance at balance, high at some finite strain, and saturate at a low value for larger strains.  Our only measurable \(p = 3\) state, at \(\nu = \frac{7}{5}\), shows a high-low-high-low resistance oscillation as expected.  

To quantitatively compare our results to earlier IQHE results in AlAs , we review the concept of ``valley susceptibility,'' denoted by \(\chi_{v}\), which is a measure of the rate of change of valley population imbalance with applied strain \cite{gunawanPRL2006}.  It is useful to discuss \(\chi_{v}\) in analogy to the more familiar spin susceptibility \(\chi_{s} = \frac{d\Delta n}{dB} = \frac{1}{2} g^* \mu_{B} \rho \), where \(\Delta n = n_{\uparrow} - n_{\downarrow}\) is the spin population difference and \(\rho = m^* / \pi \hbar^2\) is the density of states at the Fermi level.  Similarly, \(\chi_{v} \equiv \frac{d\Delta n}{d\epsilon} = \rho E^{*}_{2}\), where \(\Delta n = n_{[100]} - n_{[010]}\) is the valley population difference.  By substituting in the expression for \(\rho\), we find that \(\chi_{v} = (1/\pi \hbar ^{2}) E^{*}_{2} m^*\).  Note that \(\chi_{s} \propto g^* m^*\) and \(\chi_{v} \propto E^{*}_{2} m^*\) \cite{footnote1, footnote2}.

\begin{figure}
\centering
\includegraphics[width=85mm]{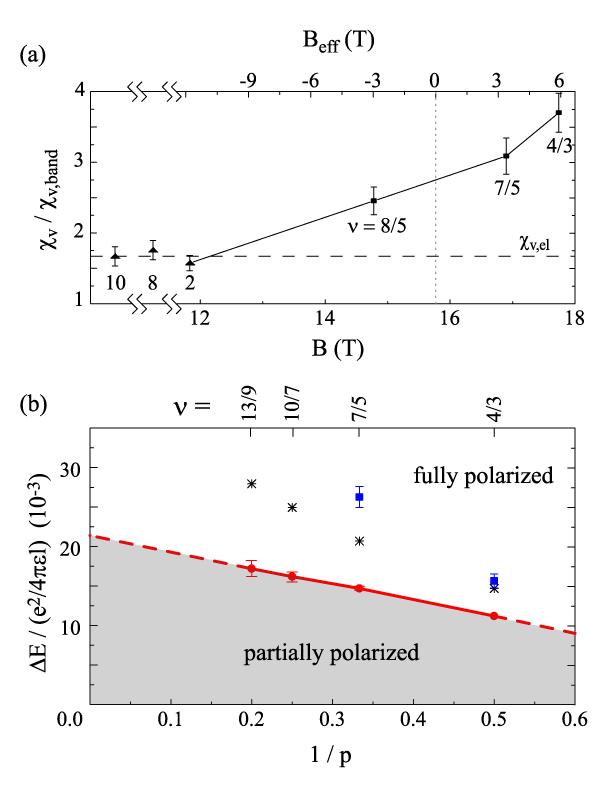}
\caption{(a) Valley susceptibility \(\chi_{v}\), normalized to its band value, for various integer (triangles) and fractional (squares) filling factors.  The horizontal dashed line represents \(\chi_{v}\) for electrons measured from low field IQHE minima (\(\nu = 2,8,10\)). (b) Ratio of polarization energy (\(\Delta E\)) to Coulomb energy (\(e^2/4\pi \varepsilon \ell\)) for various FQH states, vs. inverse CF filling factor.  The polarization energy is the amount of splitting (valley or spin) necessary to fully polarize the system.  Ratios for AlAs valley polarization are shown as blue squares, measurements of GaAs spin polarization (Ref. \cite{duPRL1995}) as black stars, and calculations of spin polarization (Ref. \cite{parkPRL1998}) as red circles.  Note that the calculations are for states around \(\nu = \frac{1}{2}\), which are equivalent to states around \(\nu = \frac{3}{2}\).  The dashed line represents the \textit{theoretical} boundary between the partially and fully polarized phases of the FQHE states.  Note that the measured polarization energies for both valley and spin systems are higher than the calculated values.}
\end{figure}

We determine \(E^{*}_{2} m^*\) (and thus \(\chi_{v}\)) using the location of our piezoresistance extrema (Figs. 3(b) and (c)).  These resistance extrema occur periodically in applied strain, when \(\Delta E_{v}\) is a multiple of \(\hbar \omega_{c}\): maxima for coincidence, minima away from coincidence.  From this we determine \(\chi_{v} = 4eB_{\nu}/h\Delta \epsilon\), where \(B_{\nu}\) is the magnetic field of the QHE state and \(\Delta \epsilon\) is the coincidence periodicity in applied strain.  From the periods of the piezoresistance traces shown in Figs. 3(b) and (c), we determine \(\chi_{v, el}\) and \(\chi_{v, CF}\) for the various QHE states, shown in Fig. 4(a).  These are plotted as a function of both \(B\) and \(B_{eff}\).  Data from IQHE states, \textit{e.g.}, at \(\nu = 2, 8,\) and \(10\), show a filling factor independent enhancement of \(\chi_{v, el}\) equal to 1.7 times the band value, represented by the horizontal dashed line in Fig. 4(a), in good agreement with the results of Ref. \cite{gunawanPRL2006}.  In contrast, \(\chi_{v, CF}\) exhibits a strong filling factor dependence, increasing to nearly four times the band value for the \(\nu = \frac{4}{3}\) state, a result not explained by our simple CF LL model.

Since there are no calculations for CFs with valley degree of freedom, we compare our data to earlier work done on GaAs 2DESs for CFs with a \textit{spin} degree of freedom.  Motivated by the measurements by Du \textit{et al.} \cite{duPRL1995} of CFs with spin in GaAs 2DESs, Park and Jain \cite{parkPRL1998} calculated a spin polarization phase diagram for the FQH states around \(\nu = \frac{1}{2}\), which are equivalent to those around \(\nu = \frac{3}{2}\).  The spin polarization energies for the various FQH states were calculated directly from the CF wavefunctions, and assumed that the CF \(g\)-factor is unchanged from the band value for GaAs electrons.  Since CFs are a product of the electron-electron Coulomb interaction, it is natural to normalize the splitting energy to the Coulomb energy for each state, \(e^2/4\pi \varepsilon \ell\), where \(\varepsilon\) is the dielectric constant and \(\ell = (\hbar /eB)^{1/2}\) is the magnetic length.  We show the results of spin measurements and calculations in Fig. 4(b), where the polarization phase boundary, normalized to the Coulomb energy, is plotted vs. inverse CF filling factor \cite{parkPRL1998}.  If we extend this analysis to our measurement, using the valley splitting in place of the spin splitting energy and assuming the band value of the deformation potential, we find surprising consistency between our results and those for CFs with spin.  The location of the valley polarization phase boundary differs from the theoretical predictions for spin by less than a factor of two, supporting the broad applicability of the CF model, and the two experimental data sets differ by only about 20\%, suggesting the near equivalence of the spin and valley degrees of freedom.

In summary, we measured the valley polarization and susceptibility for CFs around \(\nu = \frac{3}{2}\).  The CF valley susceptibility is strongly enhanced relative to the band and electron values, and exhibits strong filling factor dependence.  Plotted in a CF polarization vs. filling factor phase diagram, the data are semi-quantitatively consistent with calculations and measurements of the spin polarization of CFs in GaAs.  The results underscore the similarity of spin and valley degrees of freedom for CFs.

We thank the NSF for financial support and O. Gunawan, T. Gokmen, and J.K. Jain for discussions.  Part of our work was performed at the Florida NHMFL also supported by the NSF; we thank E. Palm, T. Murphy, J. Jaroszynski, S. Hannahs, and G. Jones for assistance.

\break


\begin{thebibliography}{20}
\bibitem[{foo({\natexlab{d}})}]{footnote4}
\bibinfo{note}{Figure 3(d) assumes that the Zeeman splitting for CFs is larger
  than the cyclotron energy at this value of \(B\). Consistent with this
  assumption, piezoresistance data for \(\nu = \frac{8}{5}\) in tilted magnetic
  fields (not shown) show no change with increasing Zeeman splitting.}
\end{thebibliography}
\end{document}